\documentclass[twocolumn,3p,times,procedia]{elsarticle}
\pdfoutput=1
\usepackage{ecrc}
\usepackage{graphicx}
\usepackage{titlesec}
\usepackage{amsmath}

\volume{00}

\firstpage{1}

\runauth{}

\jid{procs}

\CopyrightLine{2021}{Published by Elsevier Ltd.}

\usepackage{amssymb}
\usepackage{amsmath}
\usepackage{multicol}
\usepackage{amsthm}

\usepackage[figuresright]{rotating}
\usepackage{bm}
\usepackage{caption}

\usepackage{fancyhdr}

\fancyheadoffset[RO,EL]{0pt}
\usepackage{geometry}

\begin{document}

\begin{frontmatter}




\dochead{}
\title{
\begin{flushleft}
{\bf \Huge Graph Neural Network-Based Scheduling for Multi-UAV-Enabled Communications in D2D Networks} 
\end{flushleft}
}
 %

\author[]{\bf \Large \leftline {Pei Li$^a$, Lingyi Wang$^a$, Wei Wu$^*$$^a$, Fuhui Zhou$^b$, Baoyun Wang$^a$, Qihui Wu$^b$}}

\address{\bf  \leftline {$^a$Nanjing University of Posts and Telecommunications, Nanjing, 210023, China, }

\bf  \leftline {$^b$Nanjing University of Aeronautics and Astronautics, Nanjing 210016, China.}

}

\cortext[]{Wei Wu (Corresponding author)(weiwu@njupt.edu.cn).}

\fntext[]{Email address: Pei Li (2017010211@njupt.edu.cn); Lingyi Wang (b20070316@njupt.edu.cn); Wei Wu (weiwu@njupt.edu.cn); Fuhui Zhou (zhoufuhui@ieee.org); Baoyun Wang (bywang@njupt.edu.cn); Qihui Wu (wuqihui2014@sina.com).}

\begin{abstract}

In this paper, we jointly design the power control and position dispatch for Multi-unmanned aerial vehicle (UAV)-enabled communication in device-to-device (D2D) networks. Our objective is to maximize the total transmission rate of downlink users (DUs). Meanwhile, the quality of service (QoS) of all D2D users must be satisfied. We comprehensively considered the interference among D2D communications and downlink transmissions. The original problem is strongly non-convex, which requires high computational complexity for traditional optimization methods. And to make matters worse, the results are not necessarily globally optimal. In this paper, we propose a novel graph neural networks (GNN) based approach that can map the considered system into a specific graph structure and achieve the optimal solution in a low complexity manner. Particularly, we first construct a GNN-based model for the proposed network, in which the transmission links and interference links are formulated as vertexes and edges, respectively. Then, by taking the channel state information and the coordinates of ground users as the inputs, as well as the location of UAVs and the transmission power of all transmitters as outputs, we obtain the mapping from inputs to outputs through training the parameters of GNN. Simulation results verified that the way to maximize the total transmission rate of DUs can be extracted effectively via the training on samples. Moreover, it also shows that the performance of proposed GNN-based method is better than that of traditional means.


\end{abstract}


\end{frontmatter}


\section{Introduction}
Due to the fast and flexible deployment features, unmanned aerial vehicles (UAVs) have attracted extensive attention from both academia and industry \cite{survey1}. In the past few years, UAVs have been widely used in various fields such as military operations, earth and environment monitoring, disaster management, good delivery, precision agriculture, intelligent surveillance etc \cite{ChenMK1}. In the field of communications, UAVs are envisioned as a vital element of future wireless network technologies that will expand network coverage, improve deployment flexibility and increase system throughput \cite{ref1, ref2}. Compared with terrestrial base stations (BSs) whose the locations are pre-determined and fixed, UAVs can adaptively control its position to react as needed to requests for on demand services in a flexible and low-cost way \cite{ref3}.

Despite promising prospects for UAVs in many fileds like those mentioned above, some key issues must be dealt with seriously to effectively use them to realize seamless connectivity and ultra reliable communication in the future \cite{ref4}. Among them, UAV trajectory design and position deployment for sensing and communications are two important issues which have received great attention. The trajectory design mainly for the fixed-wing UAVs who are always on the move \cite{Trajectory1}, while the position deployment mainly for the rotary-wing UAVs that remain relatively stationary \cite{Position1}. For the relatively static UAVs, it may always set the position based on a pre-designed optimization algorithm. Under the assumption of quasi-static channel, such an off-line design method is effective to some extent. However, in fact, this method obviously has the defects of complex calculation, poor real-time and unrealistic. Once the environment changes, it is worth studying how to adjust the UAVs' position in time to achieve reliable communication and coverage.

D2D communication is an effective way to increase the transmission range and throughput \cite{D2D1}. Since the channel between UAV and ground user is usually assumed to be line of sight (LoS) channel, the UAV-to-ground channel will cause serious interference to D2D communications and thus compromise the system performance, how to integrate the UAV-enabled information dissemination and ground D2D transmission has attracted much attention. The authors in \cite{D2DUAV1} considered the power control of a single UAV in the D2D communication network, a centralized processing algorithm was proposed to maximize the system throughput. \cite{D2DUAV2} focused on the scenario that the D2D users act as the relay for UAV communications, by jointly optimizing user association, UAV scheduling, transmission power and UAV trajectory, the minimum security rate between UAVs was maximized.

Moreover, the authors in \cite{D2DUAV3} focused on D2D-enhanced UAV nonorthogonal multiple access (NOMA) network architecture, in which the ground users can reuse the time-frequency resources assigned to NOMA links. It proposed a novel and efficient graph-based file dispatching method with a hypergraph-based grouping algorithm.
D2D transmission was applied to extend wireless coverage of UAV in \cite{D2DUAV4}. It focused on the optimal set of UAV-served users rather than the transmission performance of D2D users. An UAV-assisted D2D network was studied in \cite{D2DUAV5}, in which a new trade off between the throughput and channel switching cost was proposed. Instead of transforming the optimization problem into a convex form, it introduced game theory and constructed a distributed framework, which greatly reduced the computational complexity and makes it more suitable for scenarios with high-speed changed channels. Random walk model was also considered, in which D2D users move continuously and UAV flies around a central point \cite{D2DUAV6}. Deep reinforcement learning was applied to solve the real-time resource allocation problem.


We note that the solutions of the above mentioned studies are not limited to convex optimization approaches. On the one hand, we know that some mathematical problems modeling of many complex network scenes are difficult to be transform into the equivalent convex form, and most of the widely used convex approximation skills can undermine optimality \cite{WuW}. On the other hand, the popularity of smart devices enable users a certain degree of mobility, which makes the acquisition of real-time channel state information even worse. As a result, some efficient and low-complexity methods such as graph theory, game theory and deep reinforcement learning have been explored and exploited \cite{ChenMK2}.

In many fields, such as communication signal processing, image processing and natural language processing, machine learning has become a common method to solve large-scale complex non-convex problems, as long as the training samples are available \cite{ML1}. Machine learning can be mainly divided into supervised learning, unsupervised learning and semi-supervised learning. The difference among them is reflected in the proportion of labeled samples \cite{ZhouY}.

Recently, graph neural networks (GNNs) have been introduced to solve large-scale wireless resource management problems \cite{main1}. Some important properties were proposed in \cite{main1}, including permutation equivalence property and the equivalence with distributed optimization algorithms, which can be used to analyze the performance and generalization of GNN-based methods. The authors in \cite{TL1} proposed an intelligent reflecting surface (IRS)-based wireless network. To obtain the channel state information (CSI) of each communication link, a deep neural network was applied to parameterize the mapping from the received pilots to the optimized system configuration. Part of the channel information that can be obtained was included in the pilot information, and the interaction between users was captured through GNN. \cite{TL3} considered a novel graph embedding based method for link scheduling in D2D networks, in which the graph embedding process was based on the distances of both communication and interference links without requiring the CSI. Two ways of learning were reflected in the simulation experiment of \cite{TL3}, i.e., supervised learning and unsupervised learning. For supervised learning, it was utilized to approximate optimal link scheduling strategy, while the unsupervised learning was employed to maximize the sum throughput of the proposed system.

From the above analysis, one may note that unsupervised learning is somewhat more popular than supervised learning. This is because that the formulated mathematical objective can be treated as the main component of the loss function, which is suitable for unsupervised learning. On the other hand, supervised learning requires a large number of labeled samples, who are not readily available since many mathematical problems of modeling are always highly non-convex and difficult to solve. Traditional convex optimization methods may obtain the optimal or near optimal solutions at the expense of high computing overhead. As a consequence, unsupervised learning becomes a better alternative.

In this paper, we focus on the deployment of relative static UAVs in D2D communication scenarios. Specifically, we consider the coexistence of the UAV-to-ground transmissions and D2D communications. Taking into account the interference of two patterns, we map the normal communication links to vertices of a graph, and map the co-channel interference links to edges of a graph. Based on the established heterogeneous graph, and taking the CSI and the coordinates of ground users as inputs, as well as taking the location of UAVs and the transmission power of all transmitters as outputs, we obtain the mapping from inputs to outputs through training the parameters of GNN. In addition, as wireless channels change with the movement of ground users, the UAVs' positions need to be constantly adjusted accordingly. According to the real-time feedback of the CSI, the positions can be adjusted through the forwarding of trained GNN rather than the high-volume recalculation, which can help reducing the flight energy consumption and the computing delay effectively.


\subsection{Contributions}
To the best of authors' knowledge, this is the first work in this field that propose GNN-based method to implement the position optimization and power control for multi-UAV-enabled communications in D2D networks. The main contributions of this paper are summarized as follows.

\begin{itemize}

\item We investigate the scenario of multi-UAV-enabled communications in D2D networks. The position arrangement and power control issue is studied to maximize the system sum throughput of UAV-to-DU links while satisfying the basic quality of service (QoS) requirement of D2D communications. A specific and efficient GNN model is proposed for the considered system and the formulated problem. We construct the transmission links and interference links as the vertices and edges, respectively. For each vertex, its feature consists of the transmission channel gain, the transmission power and the coordinate. For each edge, its feature is the interference channel gain.

\item We carry out the technical training for the formulated GNN. The iteration of each vertex is divided into two stages. In the first stage, we exploit the multi-layer perception (MLP) to aggregate the information from all the neighbors of the vertex. The aggregated information contain the features of both neighboring vertices and neighboring edges. In the second stage, by using another type of MLP, we further aggregate the output of the first stage with the feature of the vertex itself. Then, the final output is obtained with the help of a widely used activation function.

\item To demonstrate the superiority of our proposed design, we generate a large number of test samples, among which multiple D2D users and DUs are randomly distributed in a given wide range area. Numerical results show that the performance of our proposed method is significantly better than that of other benchmarks.

\end{itemize}


\subsection{Organization}
The remainder of this paper is organized as follows. Section 2 provides the system model and problem formulation. Section 3 proposes the GNN-based scheme, which includes the mapping from considered physical communication network to graph model and the training. Section 4 provides the simulation results. Finally, Section 5 concludes this paper.


\section{System Model and Problem Formulation}
\begin{figure}[!t]
\centering
\includegraphics[width=1.0\linewidth]{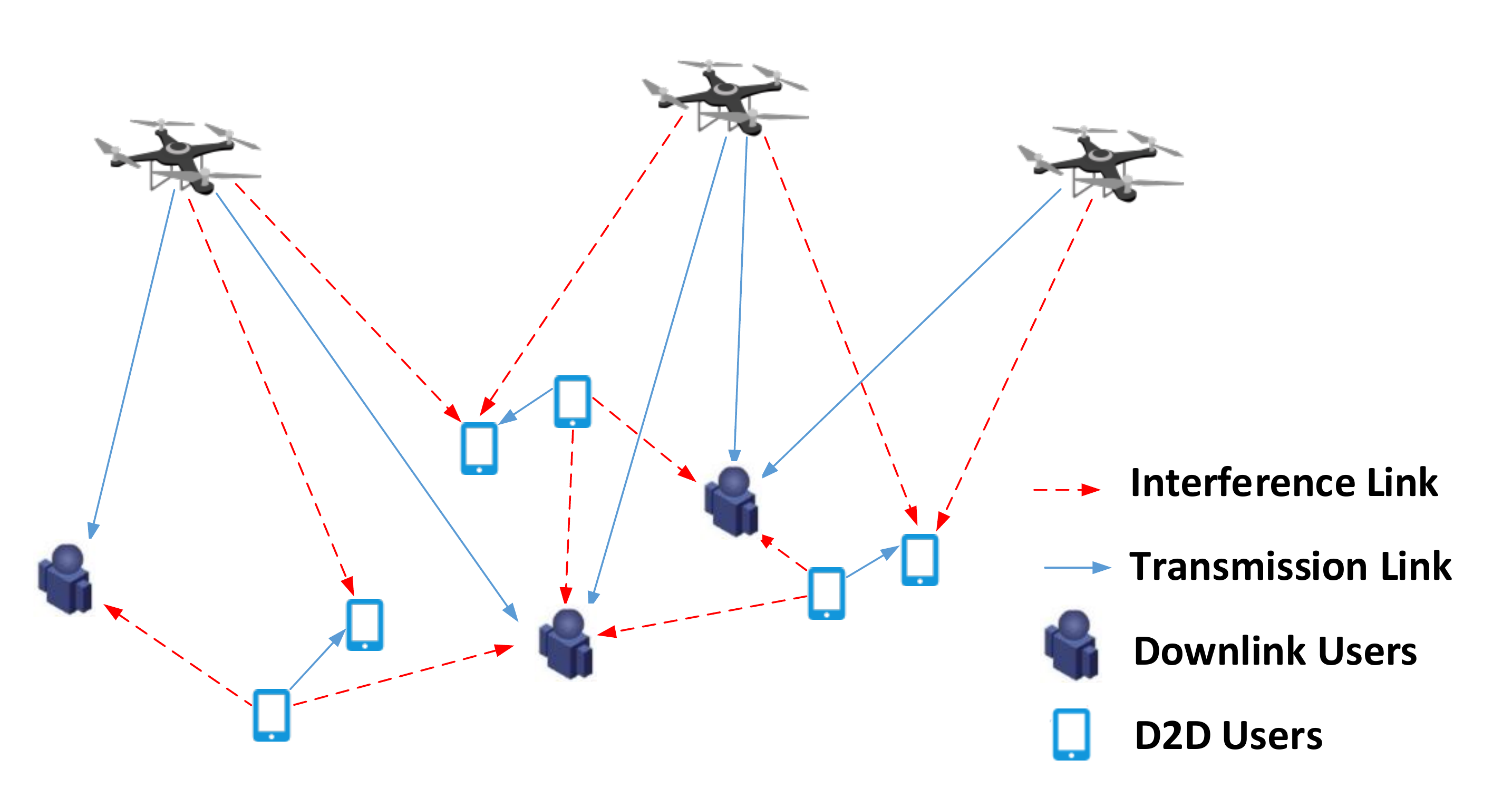}
\caption{System Model of a heterogeneous network consists of multi-UAV downlink and D2D communications.}
\label{SysMod}
\end{figure}

As shown in Fig. \ref{SysMod}, we consider a heterogeneous wireless communication system with the coexistence of multi-UAV-enabled downlink transmissions and D2D communication network, which includes $N$ UAVs, $K$ downlink users (DUs) and $M$ D2D pairs. Each D2D pair consists of one D2D transmitter (DT) and one D2D receiver (DR). Due to the limited energy storage and computing ability, multiple UAVs in the air cooperatively send the same message (such as the live high-definition video obtained from some of them) to all DUs on the ground. Meanwhile, each DT sends information to its corresponding DR. All the nodes in our system are single-antenna. It is assumed that the UAV-to-DU links share the same spectrum with D2D communications. Therefore, the co-channel interference among them must be handled carefully to achieve globally optimal system utility.


Without loss of generality, we use the three dimensional Cartesian coordinate system \cite{WangW}, with which the location of DR $m$ can be denoted as $[x_{M_m}, y_{M_m}, 0]$ and that of DU $k$ is $[x_{K_k}, y_{K_k}, 0]$, with $m \in \left\{ {1, 2,..., M} \right\}$ and $k \in \left\{ {1, 2,..., K} \right\}$. For simplicity, we assume that all UAVs are flying at the same altitude, and the location of UAV $n$ is expressed by $[x_{N_n}, y_{N_n}, H]$, where $H$ is the minimum required height to circumvent ground obstacles and $n \in \left\{ {1, 2,..., N} \right\}$.


Here, we assume that the UAV-to-DU channel is the LoS channel, and locations of all DUs are given. Then, the channel gain from UAV $n$ to DU $k$ can be expressed as \cite{ref3, WangZ}
\begin{equation}\label{channelH}
{h_{n,k}} = \frac{{{\beta _0}}}{{ {{{\left\| {{x_{N_n}} - {x_{K_k}}} \right\|}^2} + {{\left\| {{y_{N_n}} - {y_{K_k}}} \right\|}^2} + {H^2}} }},
\end{equation}
where $\beta _0$ is the channel power gain at the reference distance $d = 1$ meter.

Similarly, the channel gain from UAV $n$ to DR $m$ can be expressed as
\begin{equation}\label{channelG}
{g_{n,m}} = \frac{{{\beta _0}}}{{ {{{\left\| {{x_{N_n}} - {x_{M_m}}} \right\|}^2} + {{\left\| {{y_{N_n}} - {y_{M_m}}} \right\|}^2} + {H^2}} }}.
\end{equation}

It is worth noting that expression \eqref{channelH} and expression \eqref{channelG} have the similar structure. However, expression \eqref{channelH} refers to the transmission channel, while expression \eqref{channelG} denotes the interference channel. All air-to-ground channels are highly related to the position of UAV which is one of the main design concerns of this paper. By taking into the co-channel interference and the broadcast characteristics of the UAV-to-DU channels, the signal-to-interference and noise ratio (SINR) of DU $k$ can be given by

\begin{equation}\label{SINRk}
SIN{R_{K_k}} = \sum\limits_{n=1}^{N}\frac{{{P_{N_n}}{h_{n,k}}}}{{\sum\limits_{m = 1}^{ M} {{P_{M_m}}{g_{m,k}^u}}  + {\sigma _k}}},k \in \left\{ {1,2,...,K} \right\},
\end{equation}
where $P_{N_n}$ and $P_{M_m}$ are the transmission power of UAV $n$ and DT $m$, respectively. The Rayleigh fading channel $g_{m,k}^u = {\left( {\frac{\lambda }{{4\pi {d_1}}}} \right)^2}{\left( {\frac{{{d_1}}}{{d_{m,k}^u}}} \right)^\gamma }$ is the interference channel from DT $m$ to DU $k$ with $\lambda $ as the wavelength, ${d_1}$ as the reference distance outdoors, $\gamma $ as the path loss parameter relied on the transmission scene and $d_{m,k}^u$ as the distance between DT $m$ and DU $k$. ${\sigma _k}$ is the Gaussion noise.
Similarly, the SINR of DR $m$ is expressed as
\begin{equation}\label{SINRm}
SIN{R_{M_m}} = \frac{{{P_{M_m}}{g_{m}}}}{{\sum\limits_{i =1, i\neq m}^{ M} {{P_{M_m}}{g_{i,m}^d}}  + \sum\limits_{n =1}^{N}{P_{N_n}}{g_{n,m}} + {\sigma_m}}},
\end{equation}
for $m \in \left\{ {1, 2,..., M} \right\}$,
where ${g_m} = {\left( {\frac{\lambda }{{4\pi {d_1}}}} \right)^2}{\left( {\frac{{{d_1}}}{{{d_m}}}} \right)^\gamma }$ represents the channel between DT $m$ and DR $m$ with ${d_m}$ as the distance between them, the Rayleigh fading channel $g_{i,m}^d = {\left( {\frac{\lambda }{{4\pi {d_1}}}} \right)^2}{\left( {\frac{{{d_1}}}{{d_{i,m}^d}}} \right)^\gamma }$ is the interference channel from DT $i,\;\left( {i \ne m} \right)$ to DR $m$. ${\sigma_m}$ is the Gaussion noise. It is worth noting that the interference received at DR $m$ contains two parts, one part is that from other D2D pairs and other part is from UAVs.

Our goal is to maximize the sum throughput of DUs while guaranteeing the basic QoS requirements of all the D2D pairs.
In summary, the problem can be formulated as \cite{ref3, WuW}
\begin{equation}\label{OriginalP1}
\begin{aligned}
\mathop {\max }\limits_{{x_{N_n}},{y_{N_n}},{P_{M_m}},{P_{N_n}}} & {\sum\limits_{k = 1}^{K} {{{\log }_2}\left( {1 + SIN{R_{K_k}}} \right)} }\\
s.t.\ &0 \le {P_{N_n}} \le P_{\max }^u,\forall n \in \left\{ {1,2,...,N} \right\},\\
{\rm{        }}&0 \le {P_{M_m}} \le P_{\max }^d,\forall m \in \left\{ {1,2,...,M} \right\},\\
{\rm{        }}& {\log _2}\left( {1 + {{SIN}}{{{R}}_{M_m}}} \right) \ge {R_{\min }},\\
& \ \ \ \ \ \ \ \ \ \ \forall m \in \left\{ {1,2,...,M} \right\},
\end{aligned}
\end{equation}
where $R_{\min}$ is the required minimum transmission rate of each D2D pair.

The first two constraints in \eqref{OriginalP1} refer to the power consumption of UAVs and DTs, which cannot exceed their tolerance limit. The third constraint indicates that the interference from the UAVs and that of other D2D transmitters must ensure that the transmission rate of each D2D link cannot be lower than the threshold ${R_{\min }}$. It can be found that problem \eqref{OriginalP1} is non-convex and difficult to solve directly, since the variables are highly coupled in the objectives and constraints. In the next section, instead of resorting to traditional optimization skills, such as the alternating optimization, block coordinate descent, etc, we will provide the details of GNN-based approach to solve this problem optimally and efficiently.

\section{Graph Neural Network}
In this section, we first convert the proposed multi-UAV D2D communication system into a graph model, and then provide the detailed training process.

\subsection{Relational GNN model}

\begin{figure}[htbp]
\centering
\includegraphics[width=0.5\linewidth]{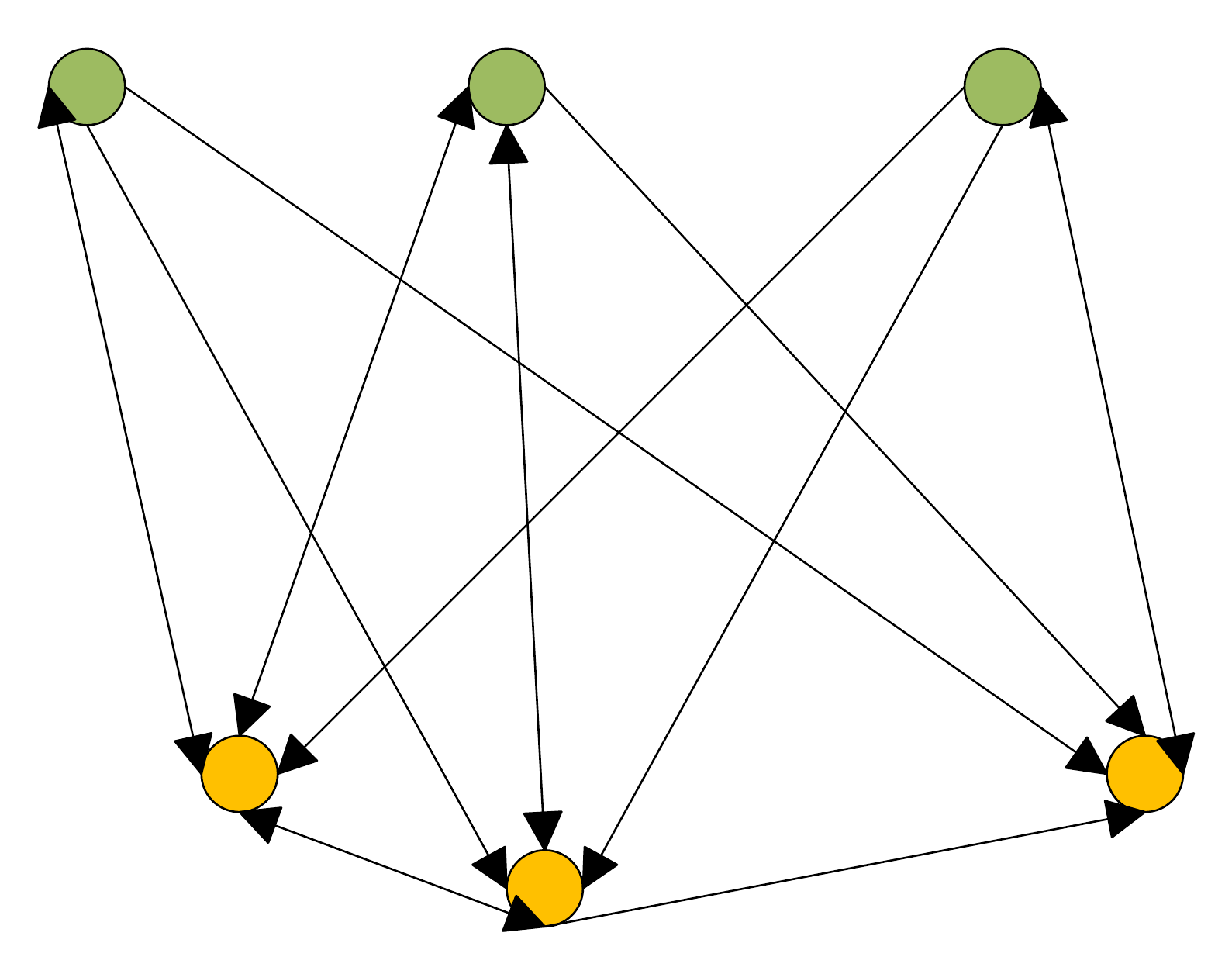}
\caption{Relational GNN model with single UAV.}
\label{oneUAV}
\end{figure}

In order to facilitate a better understanding, in this subsection, we first show the construction of the relational GNN model with single UAV. Then, based on this, we further demonstrate the GNN model with multiple UAVs, in which the relation will be more complex. For the single UAV case, as shown in Fig. \ref{oneUAV}, each UAV-to-DU link is treated as a node and is circled in green, meanwhile, each D2D link is also treated as a node and is circled in yellow. We have multiple green nodes corresponding to multiple DUs and have many yellow nodes corresponding to different D2D pairs. The directed edge denotes the interference among UAV-to-DU link and D2D link. In particular, which node the arrow points to indicates the transmission at that node is interfered with by the communication at another node on the directed edge. The double-side arrow means that the transmission at both ends of the arrow leads to mutual interference. There is no interference among green nodes due to the broadcast of the same message.

Thus, we formulate the relational GNN model of the proposed system as a set $G{\rm{ = }}\left\{ {V,E,s,t} \right\}$, where $V = \left\{ {{v_1},{v_2},...,{v_{\left| V \right|}}} \right\}$ is the vertex set, and $E$ is the edge set. $s: V \mapsto {\mathbb{C}^{{e_1}}}$ and  $t: E \mapsto {\mathbb{C}^{{e_2}}}$ represent mapping both vertices and edges to their corresponding features, respectively, where ${e_1}$ is the feature dimension of vertex and ${e_2}$ is the feature dimension of edge. In regard to the values of ${e_1}$ and ${e_2}$, let's take this step further. For ${e_1}$, it is extracted from the features of transmission power, channel gain and coordinate of both the green and yellow nodes. Since the coordinate contains both the horizontal and vertical axis variables, so we have the value ${e_1}=4$. For ${e_2}$, it is extracted from the features of interference channel. Since the ground-to-ground channels are given with fixed distance and the coordinate values have been extracted above, so we have the only feature which is the channel gain $\beta {}_0$ and have the value ${e_2}=1$.


\begin{figure}[htbp]
\centering
\includegraphics[width=0.4\linewidth]{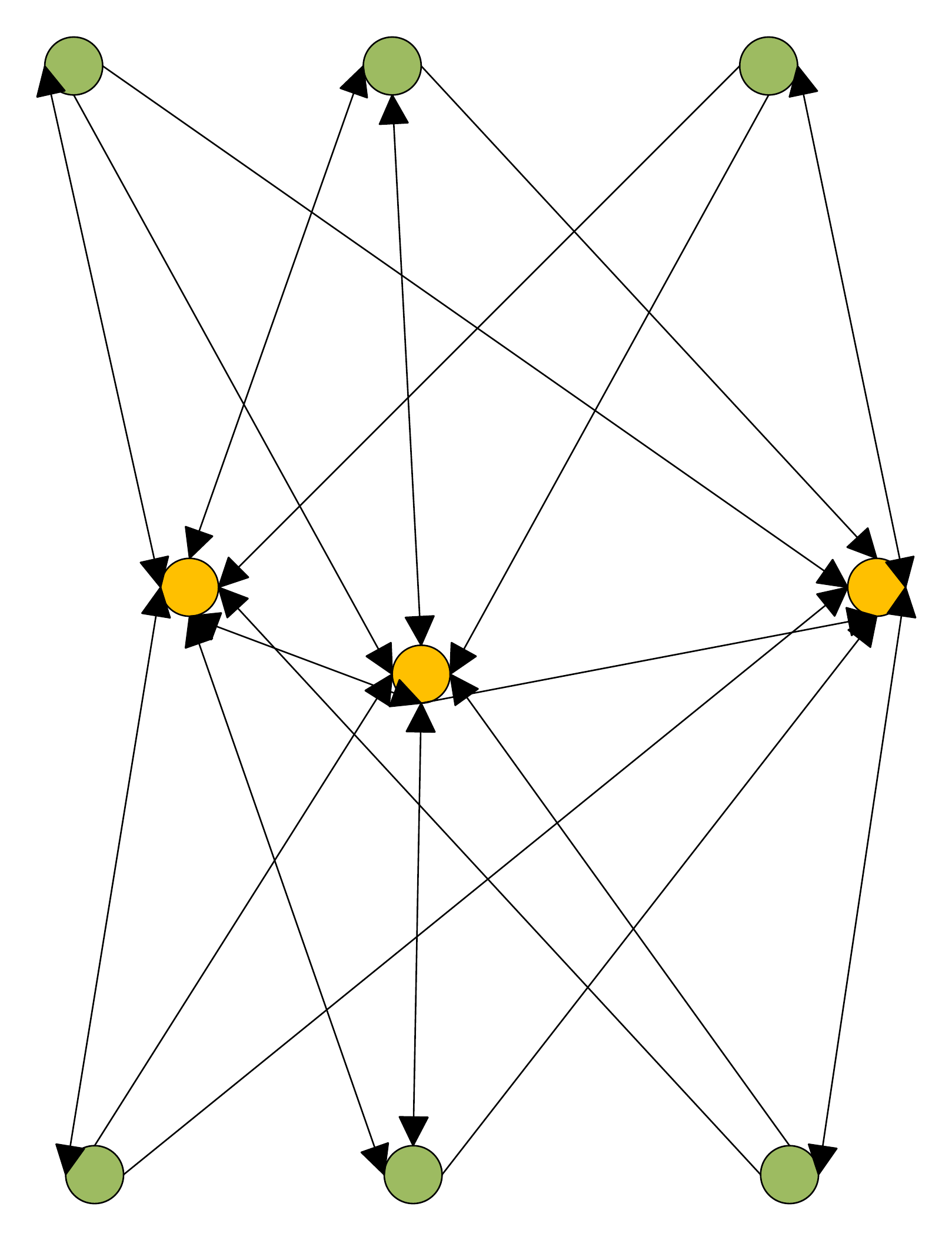}
\caption{Relational GNN model with multiple UAVs.}
\label{2UAVs}
\end{figure}

Based on the results in Fig. \ref{oneUAV}, we further extend the case to multiple UAVs. For display purposes, a graph topology corresponding to two UAVs is given in Fig. \ref{2UAVs}. From Fig. \ref{2UAVs}, it can be seen that the topology has some degree of upper and lower symmetrical structure, where the top green nodes sharing one UAV and the bottom green nodes sharing the other one. Such a symmetry remains when the number of UAVs increases. Compared with the graph structure of one single UAV, the number of nodes in the multi-UAV case will be increased from $K+M$ to $NK+M$. There is still no interference among different UAV-to-DU links.

According to the above discussions, we define the element of adjacency feature tensor ${\bf{A}} \in {\mathbb{C}^{ (NK+M)   \times (NK+M) \times 1}}$ as
\begin{equation}\label{Adjacency}
\begin{aligned}
{\bf{A}}_{i,j} = \left\{ {\begin{array}{*{20}{c}}
{{\beta _0}}, &{{\rm{if }} \ i \le NK, j > NK}, \\
{{\tilde g_{i,j}}}, &{{\rm{if }} \ i > NK, \left\{ {i,j} \right\} \in E^d}, \\
{0}, &{{\rm{otherwise},}}
\end{array}} \right.
\end{aligned}
\end{equation}
where $E^d$ is a edge set, whose element corresponds to the interference link. The value 0 means that the two nodes are not connected in graph topology. $\tilde g_{i,j}$ denotes the interference channel gain between two D2D pairs or that between a D2D pair and UAV-to-DU link. The specific definition of $\tilde g_{i,j}$ can be given as
\begin{equation}\label{ChannelDefg}
\begin{aligned}
\tilde g_{i,j} = \left\{ {\begin{array}{*{20}{c}}
{{g_{i-NK,j-NK}^d}}, &{{\rm{if }} \  j > NK}, \\
{{g_{i-NK,j \ \text{mod}\  K}^u}}, & \rm{otherwise}.
\end{array}} \right.
\end{aligned}
\end{equation}

\subsection{Training}

Considering a convolutional graph neural network with multiple convolutional layers. The convolutional layer encapsulates the hidden representation of each vertex by gathering feature information from its neighbors. After the features are aggregated, the nonlinear transformation is applied to the result output. By stacking multiple layers, the final hidden representation of each vertex will receive messages from all of the neighbors.

\begin{figure}[htbp]
\centering
\includegraphics[width=1\linewidth]{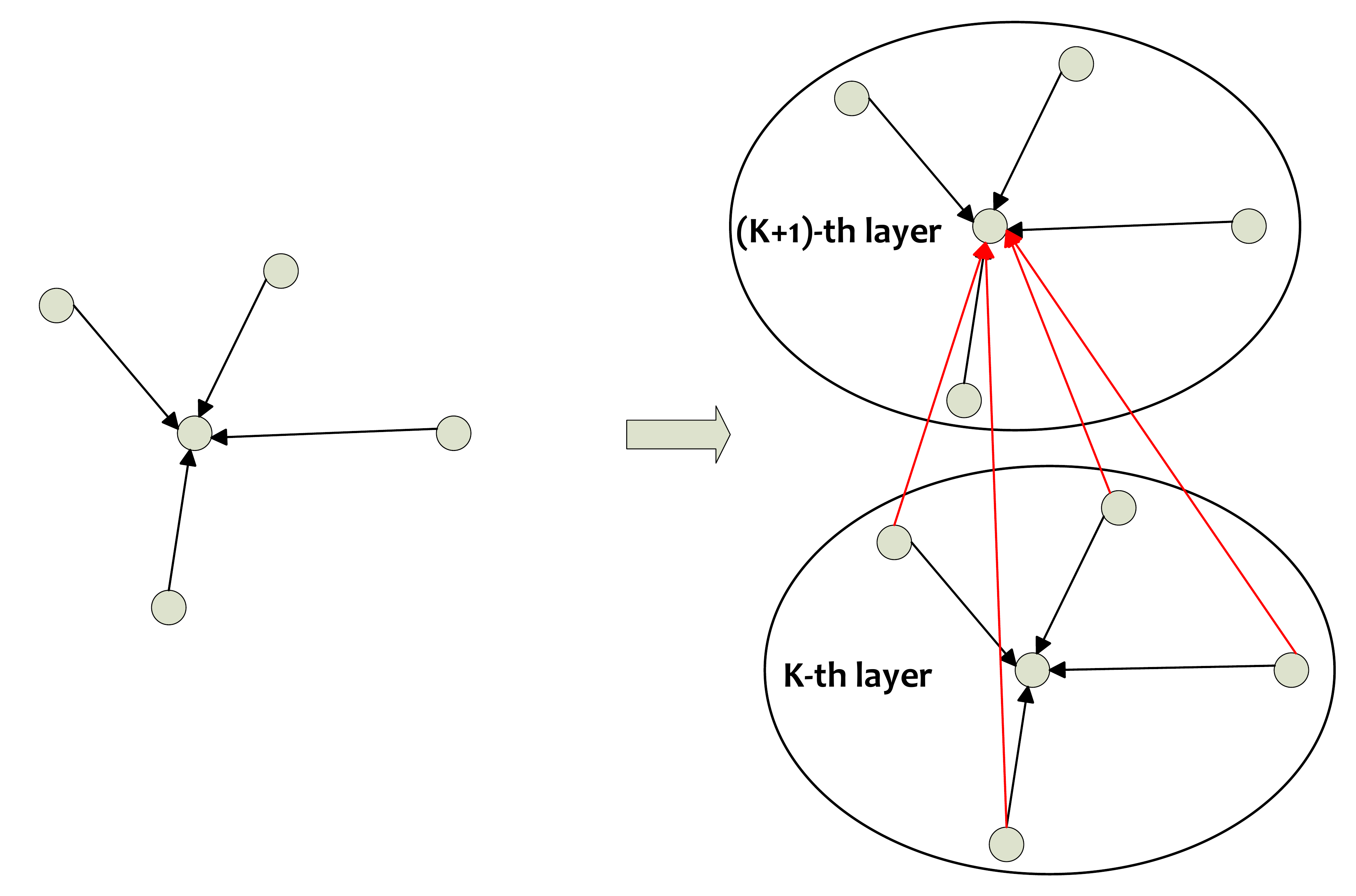}
\caption{Aggregation from the neighbors.}
\label{GCN1}
\end{figure}

The basic GNN update process is shown in Fig. \ref{GCN1}. Each vertex $i$ only aggregates the information of its neighbors $j \in \mathcal{N}_i$ during each iteration. Here, with the increase of the number of iterations, the convergence of the vertex $i$ gradually increases and eventually the training process tends to be stable. Moreover, the training process shown in Fig. \ref{GCN1} is also applicable to the case where the graph structure is a non-weighted graph, i.e., the edges of the graph involve only connectivity and carry no information. According to the definition of the adjacency matrix in \eqref{Adjacency}, the edge of our proposed graph model carries the information of interference channel. Hence, we define the update rule of the $l$-th layer at vertex $i$ as
\begin{equation}\label{forward1}
x_i^{\left( {l + 1} \right)} \leftarrow f^l\left( {x_i^{\left( l \right)},g^l\left( {x_j^{\left( l \right)},{{\bf A}_{i,j}},j \in {{{\mathcal N}}_i}} \right)} \right),
\end{equation}
where $g^l(\bullet)$ represents a set function that aggregates information from the node's neighbors, and $f^l(\bullet)$ is a function that combines aggregated information with its own information. We apply different multi-layer perceptions (MLPs) for these two functions, i.e., $\text{MLP}_1^l$ for $g^l(\bullet)$ and $\text{MLP}_2^l$ for $f^l(\bullet)$. In $\text{MLP}_1^l$, it would first aggregate information from the neighbors and their edges, and then generate the aggregated neighborhood information. The output of $\text{MLP}_1^l$ will be part of the input of $\text{MLP}_2^l$. The output of $\text{MLP}_2^l$ is processed by the activation function, usually be $\text{RELU}(\bullet)$, and the final output is obtained. Based on the above procedure, expression \eqref{forward1} can be transformed as
\begin{equation}\label{forward2}
\begin{aligned}
&y_i^{\left( {l + 1} \right)} =\text{MLP}_2^l\left( {x_i^{\left( l \right)},\text{MLP}_1^l\left( {x_j^{\left( l \right)},{{\bf A}_{i,j}},j \in {{{\mathcal N}}_i}} \right)} \right),\\
&x_i^{\left( {l + 1} \right)} = \text{RELU}\left(y_i^{\left( {l + 1} \right)}\right).
\end{aligned}
\end{equation}

\begin{figure*}[htp]
\centering
\includegraphics[width=0.8\linewidth]{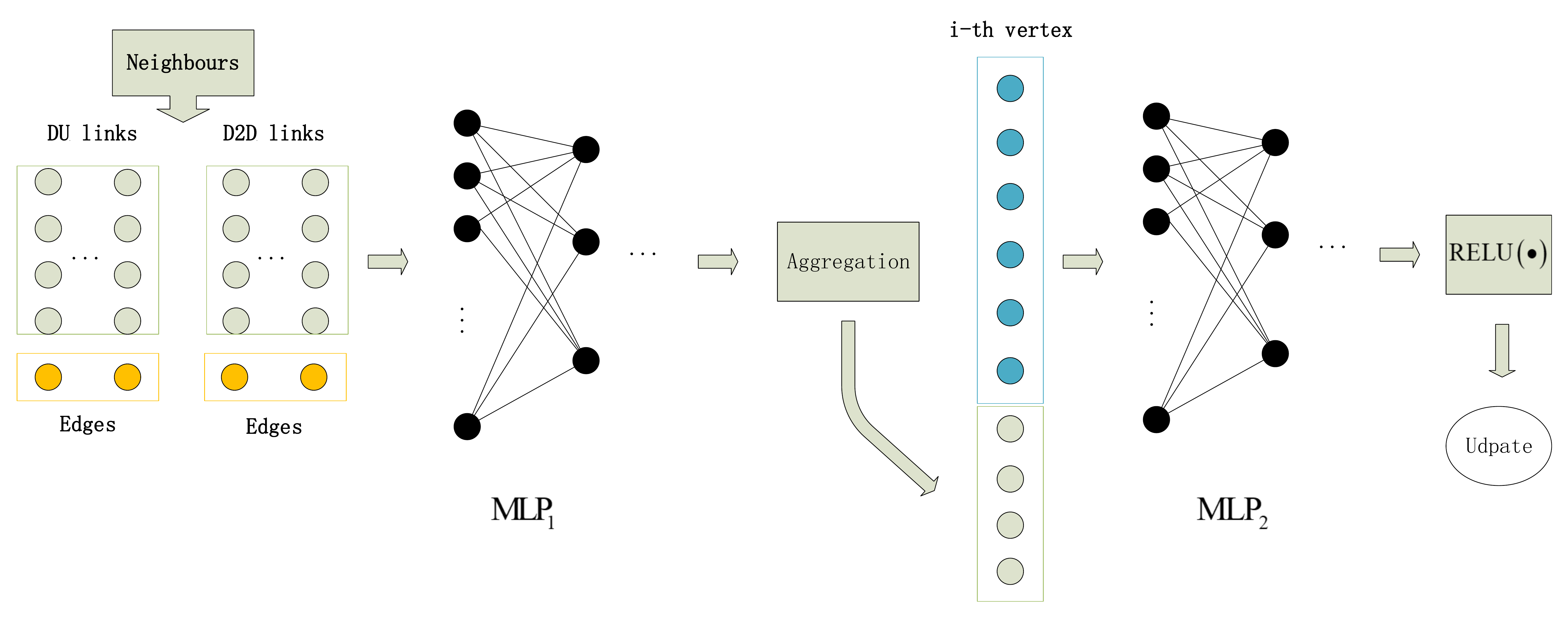}
\caption{An illustration of the update rule for the proposed GNN-based model.}
\label{UAVs}
\end{figure*}

Fig. \ref{UAVs} takes an illustration of the update rule for the proposed GNN-based model. The gray dot on the left represents the information of neighbor vertex, and the yellow dot represents the information of the edge corresponding to each neighbor vertex. The information of all the vertices and edges are spliced and input into $\text{MLP}_1^l$, then the output of $\text{MLP}_1^l$ and the information of the vertices themselves are aggregated (shown as a vector splicing in the proposed system) as the input of $\text{MLP}_2^l$. The output of $\text{MLP}_2^l$ processes through the $\text{RELU}(\bullet)$ function. That's a round of information update of the vertex.

Since the original problem \eqref{OriginalP1} is highly non-convex and difficult to solve directly, and it is impossible to obtain samples with accurate labels, we consider setting the loss function to update the network parameters.

It is worth noting that \eqref{OriginalP1} is an objective maximization problem. To minimize the loss, based on the objective function, we have the following loss function.
\begin{equation}\label{loss1}
l =  - {\sum\limits_{k = 1}^{K} {{{\log }_2}\left( {1 + SIN{R_{K_k}}} \right)} }.
\end{equation}

Further, by adding constraints in \eqref{OriginalP1} to the loss function (\ref{loss1}) as a penalty, we have
\begin{equation}\label{loss2}
l =  - {\sum\limits_{k = 1}^{K} {{{\log }_2}\left( {1 + SIN{R_{K_k}}} \right)} }  + \alpha F\left(  \bullet  \right),
\end{equation}
where $\alpha$ is the penalty coefficient, which can be set from zero to infinity. $F\left(  \bullet  \right)$ is a penalty item, whose expression is given as
\begin{equation}\label{penalty}
\begin{aligned}
F\left(  \bullet  \right) = \sum\limits_{n = 1}^N {\left( {{P_{N_n}} - P_{\max }^u} \right)^+}  + {\sum\limits_{m = 1}^M {\left( {{P_{M_m}} - P_{\max }^d} \right)^+}  + } \\
 \sum\limits_{m=1}^{M}\left(R_{\min}-{\log _2}\left( {1 + {{SIN}}{{{R}}_{M_m}}} \right)\right)^+.
\end{aligned}
\end{equation}

There are a large number of items in \eqref{penalty}, it will greatly reduce the training speed, so we need to simplify it accordingly. Specifically, \eqref{penalty} mainly includes two parts, one is transmission power, and the other is data rate. For transmission power ${P_{{N_n}}}$, through introducing Sigmoid operation, it can be re-expressed as ${P_{{N_n}}} = P_{\max }^u{\rm{sigmoid}}\left( {{P_{{N_n}}}} \right)$. Similarly, we have ${P_{{M_m}}}$ as ${P_{{M_m}}} = P_{\max }^d{\rm{sigmoid}}\left( {{P_{{M_m}}}} \right)$. By substituting them into \eqref{penalty}, the expressions ${\left( {{P_{{N_n}}} - P_{\max }^u} \right)^ + }$ and ${\left( {{P_{{M_m}}} - P_{\max }^d} \right)^ + }$ are always be equal to zero for every $m$ and $n$, so the transmission power related terms can be removed from \eqref{penalty}. As a result, the expression \eqref{loss2} can be simplified as

\begin{equation}\label{loss3}
\begin{aligned}
l =  - {\sum\limits_{k = 1}^{K} {{{\log }_2}\left( {1 + SIN{R_{K_k}}} \right)} } \\
+ \alpha  \sum\limits_{m=1}^{M}\left(R_{\min}-{\log _2}\left( {1 + {{SIN}}{{{R}}_{M_m}}} \right)\right)^+.
\end{aligned}
\end{equation}

The proposed GCN-based method employs a loss function \eqref{loss3} for the last layer to perform the parameter update. It is a regular form of unsupervised learning. In the next section, we will implement and validate the proposed scheme by simulation experiments.

\section{Numerical Results}
In this section, we provide numerical results to validate the effectiveness of the GNN-based scheme. Here, we introduce three schemes for comparison, including 1) alternate optimization (AO) scheme for the original problem, 2) random deployment of UAVs' position, and 3) fixed power allocation. Some of the main parameters are listed in Table \ref{table1}.

The graph neural network adopted includes one input layer, three hidden layers and one output layer. The number of parameters of the hidden layer is $\left\{ {32,64,32} \right\}$. In the training process, we take each user's coordinates and CSI as input, and the each UAV's power allocation and coordinates as output.

\begin{table}

\caption{simulation parameters}
\label{table1}
\begin{center}
\begin{tabular}{p{4cm}cc}

\hline

Parameter & Symbol & Value   \\ \hline

Maximum power of UAVs & $P_{\max }^u$ & 30 dBm  \\ \hline

Maximum power of D2D transmitters & $P_{\max }^d$ & 10 dBm    \\ \hline

The noise power & ${\sigma ^2}$ & -60dBm \\ \hline

The channel power gain of $d_0 = 1$ m & ${\beta _0}$ & -30dB\\ \hline

Number of DUs & $K$ & 4 \\ \hline

Number of D2D users & $M$ & 6 \\ \hline

Number of UAVs & $N$ & 4 \\ \hline

The fixed altitude of  UAVs & $H$ & 10m \\

\hline

\end{tabular}
\end{center}
\end{table}


\begin{figure}[htbp]
\centering
\includegraphics[width=0.8\linewidth]{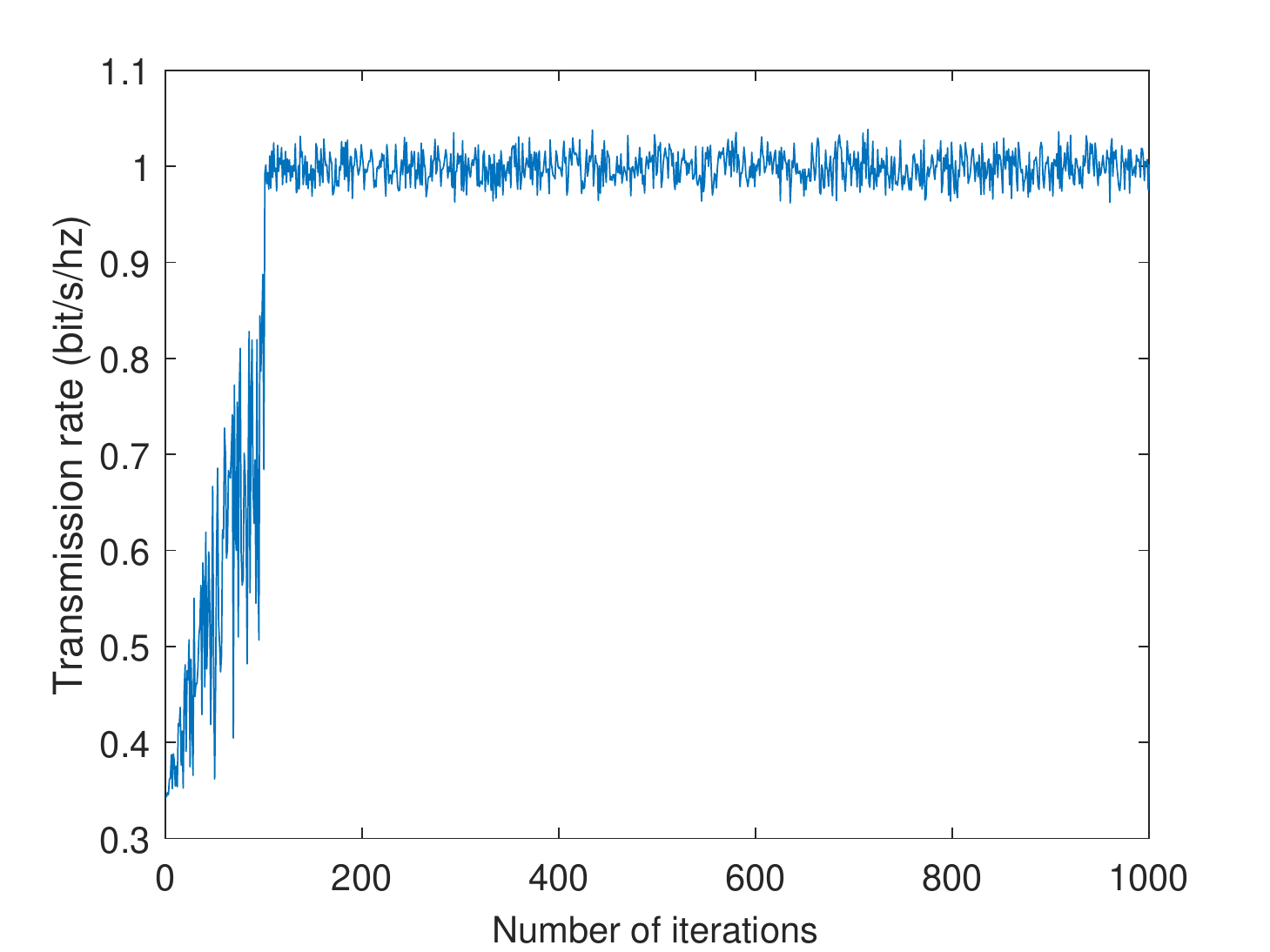}
\caption{Sum throughput of DUs versus the number of iterations.}
\label{fig0}
\end{figure}

We use program simulation to generate 500 samples as the sampling set, and generate another 200 samples as the test set. For the purpose of updating the network weights more efficient, we use the Adam optimizer \cite{Adam} instead of the classic stochastic gradient descent method. From Fig. \ref{fig0}, it can be seen that the result of the UAV-to-DU total transmission rate tends to be stable after about 100 iterations, the convergence of the proposed network is guaranteed.

\begin{figure}[htbp]
\centering
\includegraphics[width=0.8\linewidth]{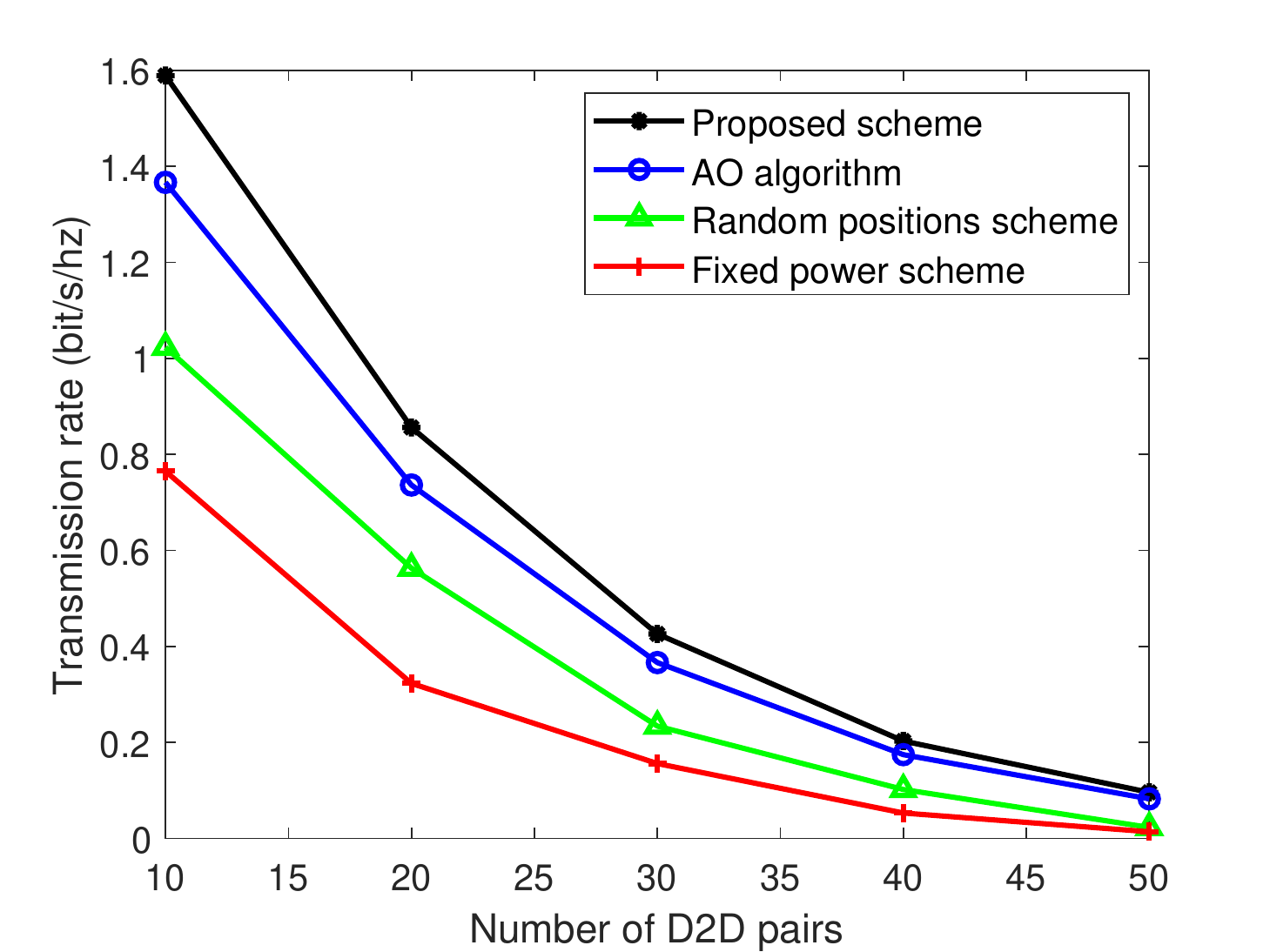}
\caption{Sum throughput of DUs versus the number of D2D pairs.}
\label{fig1}
\end{figure}

Fig. \ref{fig1} shows the the UAV-to-DU total transmission rate versus the number of D2D pairs. It reveals that our proposed algorithm has a greater improvement than other benchmark schemes in terms of downlink transmission rates. Specifically, as the number of D2D users increases, the total transmission rate of the system decreases. Since D2D users are randomly distributed in the area, the increase in the number of D2D users will not only increase the interference with DUs and DRs, but also make the deployment of UAVs more difficult. In this case, the UAV tends to fly far away from the communication area or reduce the transmission power to ensure the transmission rate of D2D users. Compared to $M=10$, when $M \ge 30$, the transmission rate will decrease by more than two thirds. Hence, in this situation, increasing the number of UAV may be a feasible way to improve the performance, as what we will discussed in the next simulation result.

\begin{figure}[htbp]
\centering
\includegraphics[width=0.8\linewidth]{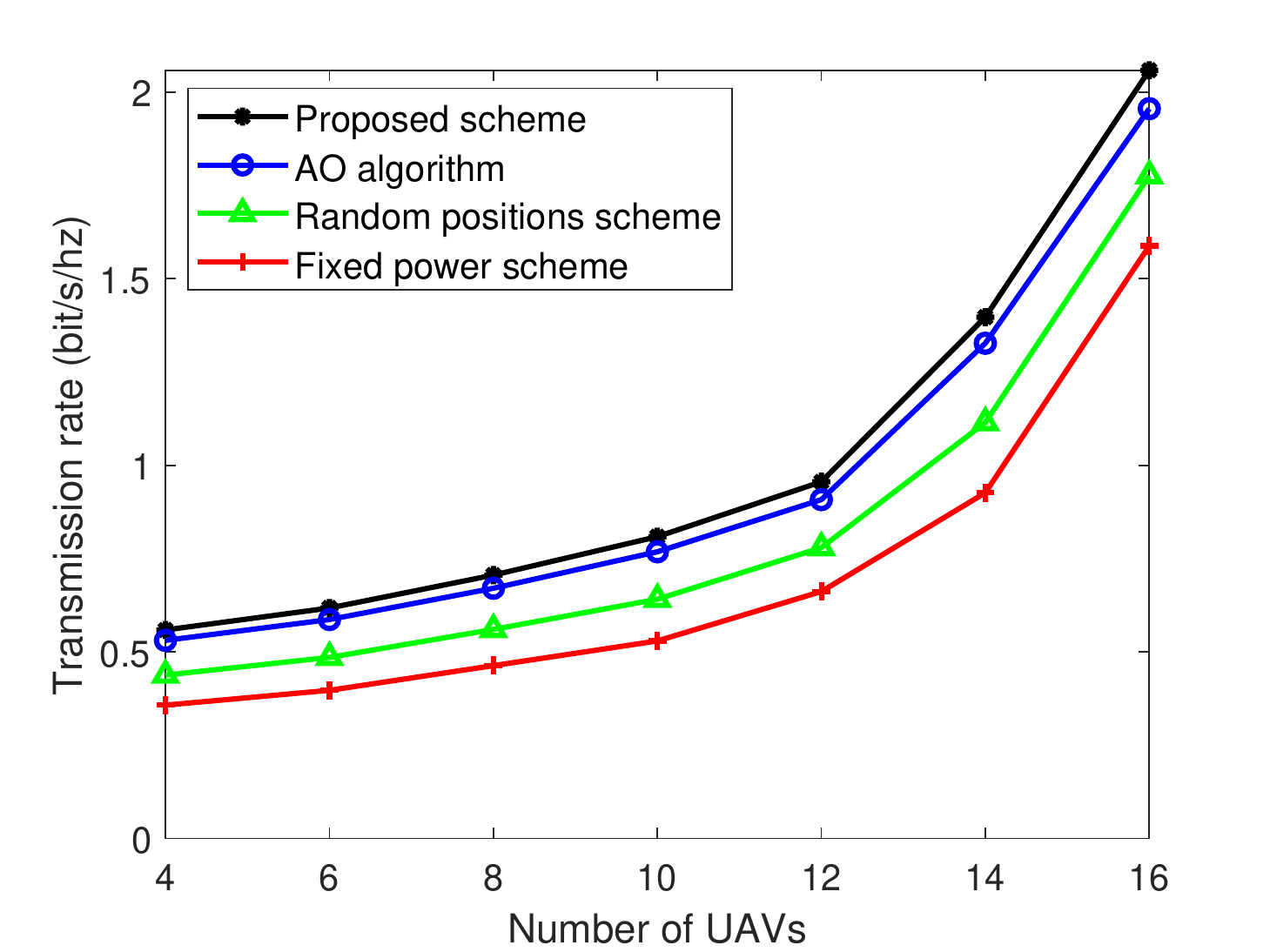}
\caption{Sum throughput of DUs versus the number of UAVs.}
\label{fig2}
\end{figure}

In Fig. \ref{fig2}, we take the total transmission rate versus the number of UAVs. It can be found that as the number of UAVs increases, the transmission rates increase significantly. This means that increase the number of UAV will not significant increase the interference with the DUs, and the interference with DRs from the newly added UAVs can be effectively suppressed by location deployment and power control. All of this reveals the obvious advantages of multi-UAV-enabled communications.

\section{Conclusion}
In this paper, the power control and position optimization problem was investigated for multi-UAV-enabled communications in D2D network. An GNN-based method was proposed to tackle the highly non-convex fractional programming problem optimally. In particular, we formulated a graph-based model for the proposed network via mapping the transmission links and interference links into vertexes and edges, respectively. Then, we trained the GNN with unsupervised learning. Finally, the trained GNN can be used to perform scheduling based on the input channel state information and the coordinates of ground users. Simulation results demonstrated the outstanding performance of UAV in facilitating the information broadcasting to ground downlink users. Moreover, the proposed GNN-based approach shows lower complexity and better system throughput than that of the benchmark scheme.

\section*{Acknowledgements}
This work was supported in part by the National Natural Science Foundation of China (61901231), in part by the National Natural Science Foundation of China (61971238), in part by the Natural Science Foundation of Jiangsu Province of China (BK20180757), in part by the open project of the Key Laboratory of Dynamic Cognitive System of Electromagnetic Spectrum Space, Ministry of Industry and Information Technology (KF20202102), in part by the China Postdoctoral Science Foundation under Grant (2020M671480), in part by the Jiangsu Planned Projects for Postdoctoral Research Funds (2020Z295).

\bibliographystyle{elsarticle-num}
\bibliography{egbib}
~~~\\
~~~\\







\end{document}